# Diagnosis of Breast Cancer Based on Modern Mammography using Hybrid Transfer Learning

**Aditya Khamparia[1], Subrato Bharati[2], Prajoy Podder[2], Deepak Gupta[3], Ashish Khanna[3], Thai Kim Phung[4], Dang N. H. Thanh[4],***

[1]*School of Computer Science and Engineering, Lovely Professional University, Punjab, India*

[2]*Institute of Information and Communication Technology, Bangladesh University of Engineering and Technology, Dhaka, Bangladesh*

[3]*Maharaja Agrasen Institute of Technology, India*

[4]*Department of Information Technology, School of Business Information Technology, University of Economics Ho Chi Minh City, Vietnam*

*\*Corresponding: Dang N. H. Thanh, thanhdnh@ueh.edu.vn*

**Abstract-** Breast cancer is a common cancer for women. Early detection of breast cancer can considerably increase the survival rate of women. This paper mainly focuses on transfer learning process to detect breast cancer. Modified VGG (MVGG), residual network, mobile network is proposed and implemented in this paper. DDSM dataset is used in this paper. Experimental results explain that our proposed hybrid transfers learning model (Fusion of MVGG16 and ImageNet) provides an accuracy of 88.3% where the number of epoch is 15. On the other hand, only modified VGG 16 architecture (MVGG 16) provides an accuracy 80.8% and MobileNet provides an accuracy of 77.2%. So, it is clearly stated that the proposed hybrid pre-trained network outperforms well compared to single architecture. This architecture can be considered as an effective tool for the radiologists in order to decrease the false negative and false positive rate. Therefore, the efficiency of mammography analysis will be improved.

# 1 Introduction

Every year, 12% of women are diagnosed with breast cancer [1]. In the US alone, 40,000 women die of breast cancer annually [2, 3]. Evidence shows that early detection of breast cancer can significantly increase the survival rate of women.

Mammography, a special X-ray of the woman's breast, is one of the most common diagnostic tools for detecting breast cancer [4, 5]. The images show masses and even calcifications, which are precursors to breast cancer. However, correctly identifying these images can be challenging for radiologists. Moreover, time constraints in assessing the images often result in incorrect diagnosis with detrimental consequences. For instance, a false negative diagnosis, that a case is normal when it is in fact an early form of breast cancer, can decrease the chance of 5-year survival significantly.

A mammogram is a type of X-ray image of the breast. Doctors use the mammogram image to identify the signs of breast cancer. So, it is capable of detecting calcifications, lumps, dimpling etc. These are the common signs showed in early stage of breast cancer. The DDSM (Digital Database for Screening Mammography) is used in this paper. DDSM is a group of labelled mammographic images. This database is maintained by the research community [6]. Figure 1 shows some types of image of DDSM.

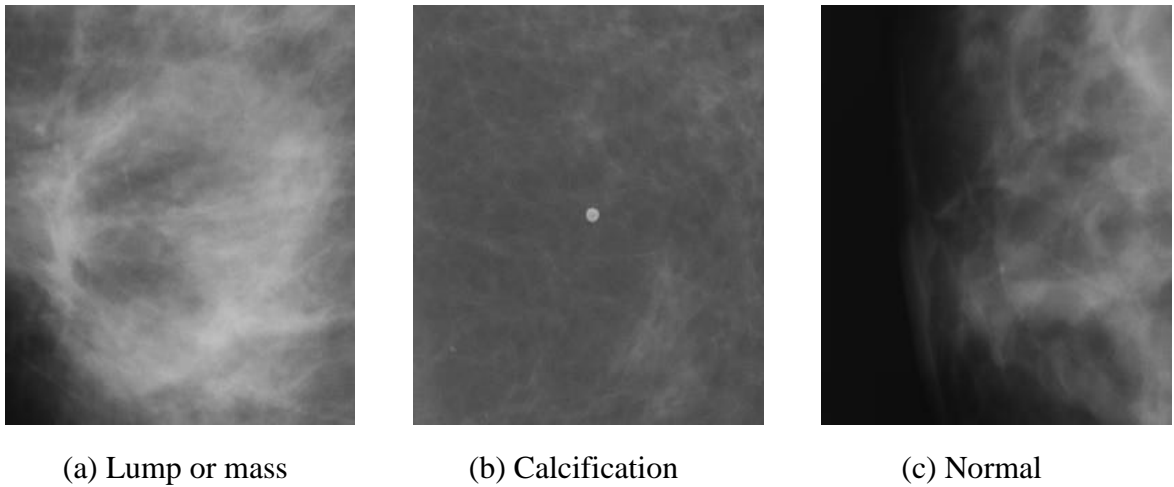

| (a) Lump or mass | (b) Calcification | (c) Normal |

**Figure 1:** Some types of image of DDSM

2,620 instances are contained in the dataset. The instances are the mammograms of patients with masses, calcifications, etc. The labels for calcification and masses are specified into four categories.

(i) benign

(ii) malignant

(iii) benign without callback

(iv) unproven.

In addition, the images have been categorized on a scale of 1-5 additionally according to the BI-RADS. BI-RADS means breast imaging, reporting and data system. BIRADS can be considered the most effective tool in order to detect the breast cancer. The score 5 shows that the mammogram results are very suspicious and the probability of breast cancer is almost 95%. In order to simply our analysis, patches are used instead of full images. It helps not only for efficient computation but also for better performance.

Because feature detection becomes easier. 10,713 patches are contained in our dataset. Table 1 summarizes the statistics of the DDSM patch data set.

**Table 1:** Summary statistics of the DDSM patch data set

|  | **Malignant** | **Benign w/o callback** | **Benign** | **Unproven** | **Total** |
|---|---|---|---|---|---|
| Calcification | 797 | 539 | 800 | 16 | 2,152 |
| Mass | 1,075 | 179 | 1,079 | 21 | 2,354 |
| Number of pathological cases | - | - | - | - | 4,506 |
| Number of non-pathological cases | - | - | - | - | 6,207 |
| Total number of patches | - | - | - | - | 10,713 |

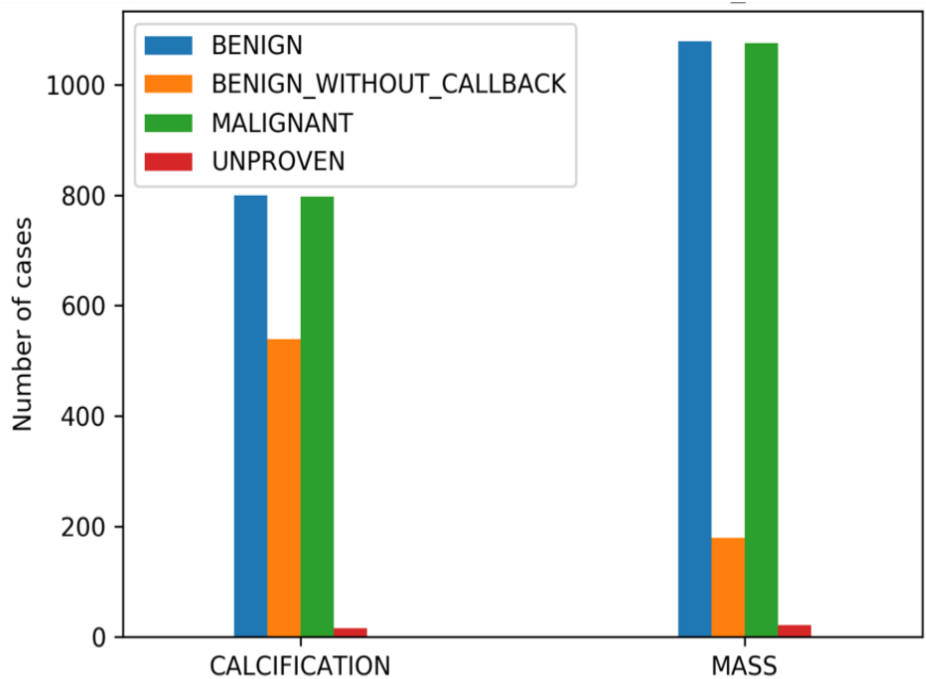

**Figure 2:** Classes and labels of the DDSM dataset

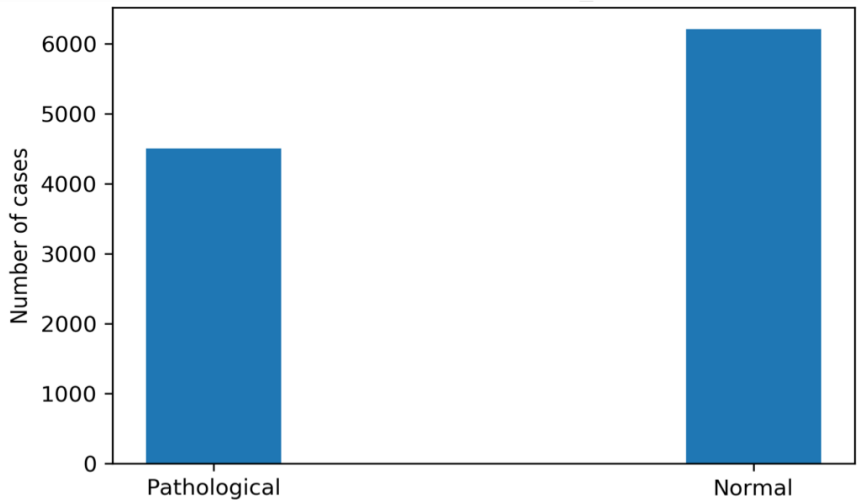

**Figure 3:** Split of the patches based on pathology

Exploring the data further yields the plots in Fig. 2 and Fig. 3. Overall, there are more cases of masses than of calcification (in Fig. 2). The number of malignant and benign cases for calcification and masses seems to be the same. For both calcification and masses, a few cases have been categorized as 'unproven'. In our analysis, we decide to mark these as pathological as it is not clear if they can be considered healthy patients or not (the number is small and should not have a strong negative impact on our predictive power). As we aim to model a binary classifier, we label all mass and calcification patches as being pathological. In total, we have 4,506 pathological and 6,027 non-pathological patches (in Fig. 3).

## 2   Literature Reviews

Deep learning-based system for classification of the images of breast tissue is proposed in [7]. For those images, 650×650 has been extracted with 400×400 pixels. Next, pre-trained VGG-16, InceptionV3 and ResNet-50 networks is conducted for the feature extraction. A 10-fold cross validation with LightGBM classifier has been conducted to the classification and extraction of deep features. That technique gets an average accuracy of 87.2% across leave on out for the purpose of breast cancer images classification [7]. In the another work [8], 4-DCNN architectures i.e. InceptionResnetV2, InceptionV3, VGG19 and InceptionV4 have been used for the classification of images of breast cancer. The size of the images is $1495 \times 1495$ of 99 pixels. Various data augmentation systems have also been developed in order to increase the accuracy. In [9], the ensemble-based architecture is proposed for multi-class image classification of breast cancer. Their conduced ensemble classifier involved; logistic regression, gradient boosting machine (GBM), majority voting to achieve the final prediction. In [10], stain-normalization technique is applied to stain images for normalization where the achieving accuracy is 87.50%. Another research [11] conducts a machine learning method where feature vectors are extracted from various characteristics i.e. texture, color, shape, etc. For the 10-fold cross validation, SVM provides 79.2% accuracy. Lastly, the paper [12] uses a fine-tuned AlexNet for automatic classification of breast cancer. They achieve the accuracy of 75.73% where patch-wise dataset is used.

Our literature review covered three topics: 1) The state-of-art deep learning architectures for the task of binary classification of images 2) Performance achieved in similar tasks as a benchmark for our algorithm 3) Studies on physicians' performance to understand the clinical implications of such algorithms.

## 2.1 State of the art architectures

VGG network is presented by the author of [13]. It is a simple model. It consists of a 13 layered CNN where 3x3 filters (fig.4) are used. VGG model has 2x2 maxpooling layers. The performance of multiple, smaller-sized kernels is comparatively better than a single larger-sized kernel. Because, the increased depth of the VGG network can support the kernel to learn more complex features.

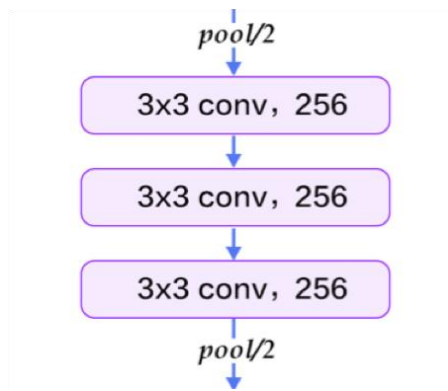

**Figure 4:** The characteristic 3x3 convolution layers of the VGG

Secondly, Residual Network (ResNet) is considered. For image classification, ResNet is the most popular architecture. It is presented by the paper [14]. Residual block can be considered a distinguishing feature in ResNet. (in Fig. 5). Residual block allows the residual network to achieve a depth of 152 layers. vanishing gradients is a common problem in DCN. This problem can be moderated by the residual block. Because of vanishing gradients, the performance of ResNet can be degraded with the increase of depth.

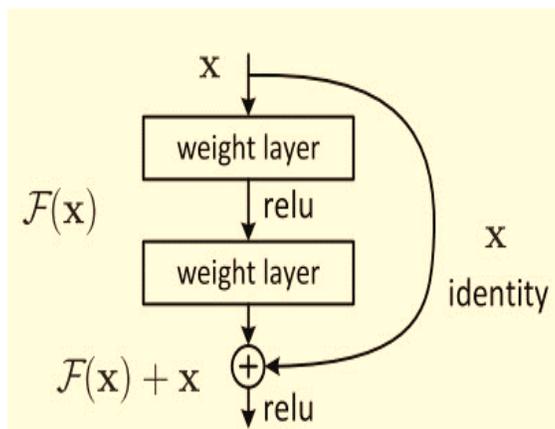

**Figure 5:** The residual model of the ResNet architecture

Finally, MobileNet is considered in this paper. MobileNet is a specialized network for mobile and embedded vision applications. This Network is introduced by Howard et al [15]. It is a basic DNN that uses depth-wise separable convolutions, which factorize standard convolutions into one depth-wise and another point-wise convolution. The advantage is that it results in far fewer parameters and helps build lighter deep neural networks that trade-off accuracy for improved latency (in Fig. 6).

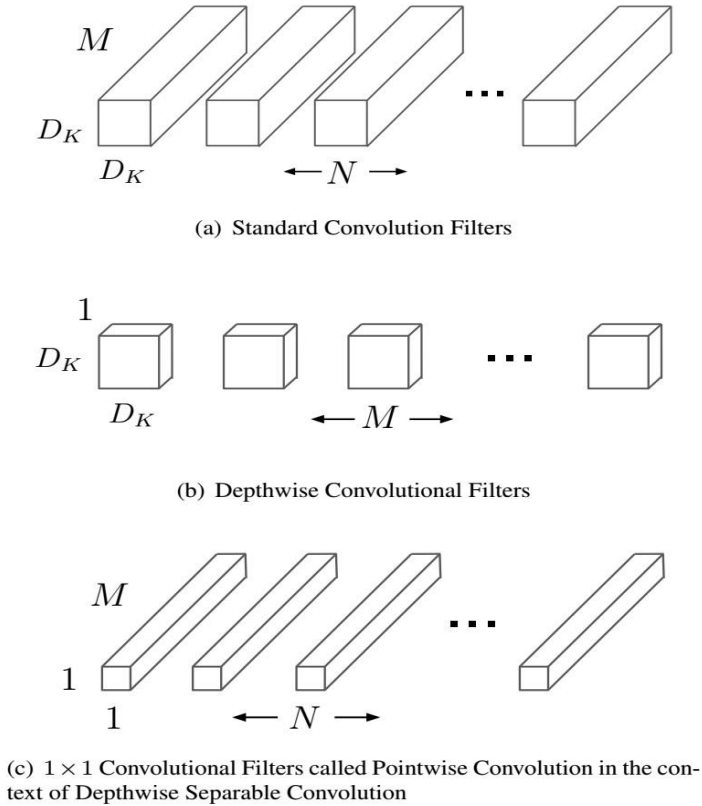

**Figure 6:** MobileNet Architecture

## 2.2 Prior art in breast cancer classification

In order to detect breast cancer, several algorithms and classification have been developed using different datasets. For instance, a paper published in 2015 obtained 85% accuracy for identifying images with a mass, and also localizing 85% of masses in mammograms with an average false positive rate per image of 0.9 [16]. Shen et al [17], developed an end-to-end training algorithm for a whole-image diagnosis. It deploys a simple convolutional design achieving a per-image AUC score of 0.88 on the DDSM dataset. We adopt this as metric as the benchmark for our algorithm, in addition to an accuracy benchmark of 85%.

## 2.3 Physician performance

Several high quality studies explored the performance of physicians diagnosing mammograms. A study by Rafferty et al looked at radiologist performance on mammographs from 1,192 patients [18]. In a first study, 312 cases (48 cancer cases) were diagnosed by 12 radiologists who recorded if an abnormality which requires a callback was present. This resulted in a sensitivity of 65.5% and a specificity of 84.1%. In a second study, 312 cases (51 cancer cases) were analyzed by 15 radiologists. They obtained additional training and also reported the the type and location of the lesion. This resulted in a sensitivity of 62.7% and a specificity of 86.2%. Another high quality study compared different diagnosis methods such as mammography, ultrasonography (US), and physical examination (PE) using a data set of 27,825 screening sessions Kolb et al. [19] and compared the results of the three diagnosis methods with the actual biopsy. The results showed a sensitivity of 77.6% and a specificity of 98.8%. However, these scores were not achieved by radiologists using only mammograms and thus do not fit well for a benchmark for this task.

Most relevant as a benchmark for our analysis is the first study by Rafferty et al [18] as the 12 radiologists restricted themselves to binary classification, which is similar to our approach.

## 2.4 Clinical significance

Medical suggestions of the mammogram diagnosis are essential in order to improve an algorithm with Clinical significance. Radiologists have a considerably higher specificity than sensitivity. So, it means that false negative rate is higher than the false positive rate. Table 2 illustrates the Comparative analysis of risks for two types (1 and 2) of diagnostic errors.

**Table 2:** Comparison of risks for type 1 and 2 diagnostic errors

| | Risks and costs of a diagnostic error |
|---|---|
| False positive | • Additional test: Costs and minimal-invasive biopsy<br>• Short-term distress/long-term risk of anxiety |
| False negative | • 5-year survival rate is strongly impacted by later detection:<br>Decreases from 93% to 72% from stage 3 to stage 2 |

False positive diagnosis indicates that the radiologist judges a normal mammogram of malignant or benign type. As a result, that patient has to visit to the clinic again and in most cases further testing through a biopsy is performed. Biopsy for breast cancer detection is minimally invasive and only a small incision is needed. However, there is a range of evidence showing the psychological effects of such false positives. According to a study from 2000, it can lead to short term distress as well as long-term anxiety Aro et al [20]. On the contrary, a false negative implies that a potentially cancerous case is misinterpreted as healthy. The consequences of this can be very severe because breast cancer, when left untreated progresses in its stages and with each stage a different 5-year life expectancy is associated (in Table 3 [21])

**Table 3:** Survival rate of breast cancer stages

| 5 years overall survival by stage | | |
|---|---|---|
| **Stage** | **5-year overall survival** | **Classification** |
| 0 | 100% | In situ |
| 1 | 100% | Cancer formed |
| 2 | 93% | Lymph nodes |
| 3 | 72% | Locally advanced |
| 4 | 22% | Metastatic |

In general, it is more important to avoid a false negative over a false positive. A mammogram is first followed by a sonography and if this is positive as well, further testing is done via a minimal invasive biopsy. However, in the future it would be great to differentiate further and decrease the false positive in the BI-RADS 3 category. As of now, 98% of patients in this category have to come back every 6, 12, 24 months for a check-up, yet do not have breast cancer. This is a large burden.

Considering Table 2 and the two interviews, we conclude that a false negative error can have more severe consequences than a false positive error. Thus, we decide to design our algorithm to have a threshold which is more sensitive than specific.

## 3 Modeling

### 3.1 Data cleaning

Before building the model, we clean the data set to convert it into the appropriate form. We assign new, binary labels to the images by categorizing all the original mass and calcification labels as 'pathological', and the normal images as 'non-pathological'. Thus, the problem is decreased to a binary classification. Next, we randomly divide the data set into train, validation, and test splits, in approximate proportions of 75:10:15 respectively. While doing so, we ensure that the splits are evenly balanced between the two classes (as evident in Fig. 7).

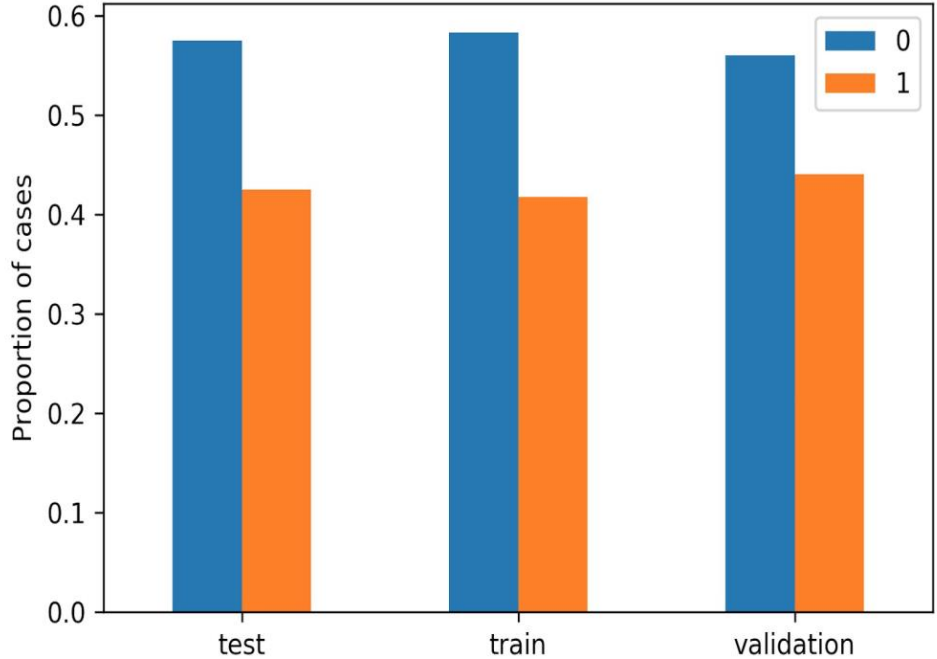

**Figure 7:** Balance of classes in the train, validation, and test splits

## 3.2 Performance evaluation

The performance is evaluated with the help of the two metrics. They are: Accuracy and AUC. AUC means area under the ROC curve. These two parameters are widely-used for evaluating classification. Accuracy means the percentage of cases that the model classifies correctly. AUC means the capability of the model to discriminate between the two classes.

## 3.3 Modeling process

The flow chart can easily describe the modeling process showed in fig. 8.

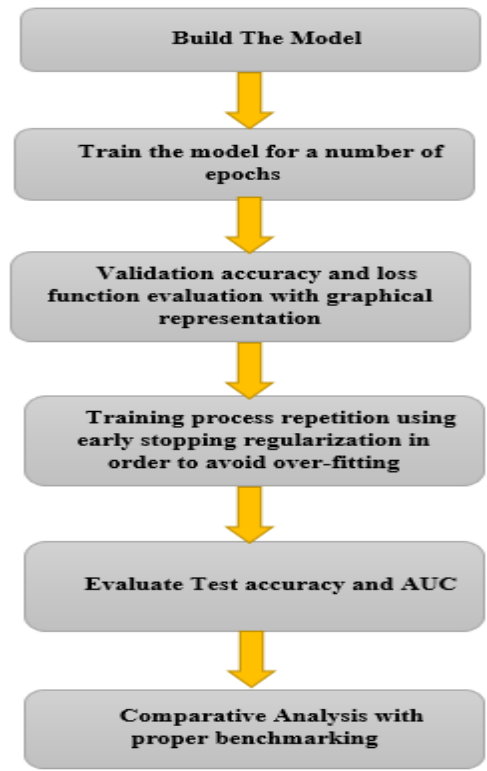

**Figure 8:** Work flow diagram of modeling process

## 3.4 Model building

Model building is the first step of model processing. It can be divided into four sub- stages. They are:

1. Construction of baseline model and performance evaluation

2. Training of popular models with various architectures, and selection of best models.
3. Deployment of regularization and data augmentation methods in order to develop the performance. Choose the best model as the final model based on the performance.

4. Tuning of hyper-parameters on the final model in order to accomplish the desired level.

**Table 4:** Architecture of the baseline model

| Layer (type) | Output Shape | Parameters |
|---|---|---|
| Conv2d_3 (2D convolution layer) | (None, 254,254,32) | 320 |
| Conv2d_4 (2D convolution layer) | (None, 252,252,64) | 18496 |
| Max_pooling2d_2 | (None, 126, 126,64) | 0 |
| Flatten_2 | (None, 1016064) | 0 |
| Dense_3 | (None, 32) | 32514080 |
| Dense_4 | (None, 1) | 33 |

Table 4 describes the details of the architecture of the baseline (simple) model. There are two 2-dimensional convolution layers. There are 32 and 64 filters. There is a dense layer in the architecture having 32 nodes on the top. At the time of the performance evaluation of the model on the test set, accuracy of 75.9% is obtained. It is about 13% lower than desired accuracy or desired benchmark.

In next stage, three image classification models are implemented. The models are VGG16, MobileNet and ResNet50. The models are then modified by tuning the feedforward, dense layers in the end to only one layer having 32 nodes. These models are originally designed to label up to 1000 classes and therefore have wide dense layers (4096 nodes).

Two additional methods have been considered after choosing the final model in order to observe if they improve performance. The two techniques are: i) Data augmentation and ii) Pre-training.

In data augmentation stage, three operations are performed on the input images. The operations are:

1) Flip the images along a horizontal axis

2) Shift vertically/horizontally within a width range of 0.2

3) Rotate randomly within a twenty-degree range.

Pre-training process includes initializing model parameters with values learned from a different data set, instead of random ones. Pre-training process not only can speed up learning but also achieve improved local optima in gradient optimization. In this paper, the best model is pre-trained using weights. ImageNet data set is trained in this case.

For the purpose of tuning on the best model, batch size and learning rates are varied to enhance the accuracy.

## 4 Results and Interpretation

### 4.1 Performance of different models

The performance of different implemented models is showed in Table 5.

**Table 5:** Performance evaluation of different models

| Model | Size of Batch | Special pre-processing | Accuracy | Number of epochs |
|---|---|---|---|---|
| Simple model | 32 | No | 75.9% | 15 |
| ResNet50 | 32 | No | 75.1% | 15 |
| Mobile Net | 32 | No | 77.2% | 15 |
| MVGG16 (Modified VGG 16) | 32 | No | 80.8% | 15 |

| MVGG16+Augmentaton | 32 | Flips, shifts, rotations | 82.8% | 30 |
| MVGG16+ImageNet (Final Model) | 32 | Pre-trained on ImageNet | **88.3%** | 15 |

At the end of the model building process, we realize that the pre-trained modified VGG16 (MVGG 16) model outperforms all others in terms of accuracy. The architecture of the model is shown in Fig. 9. It produces an accuracy of 86.9% on the test set and an AUC of 0.933. This is better than our benchmark on both metrics. In Fig. 10(b) we can see that the model starts to strongly over fit after 6 epochs.

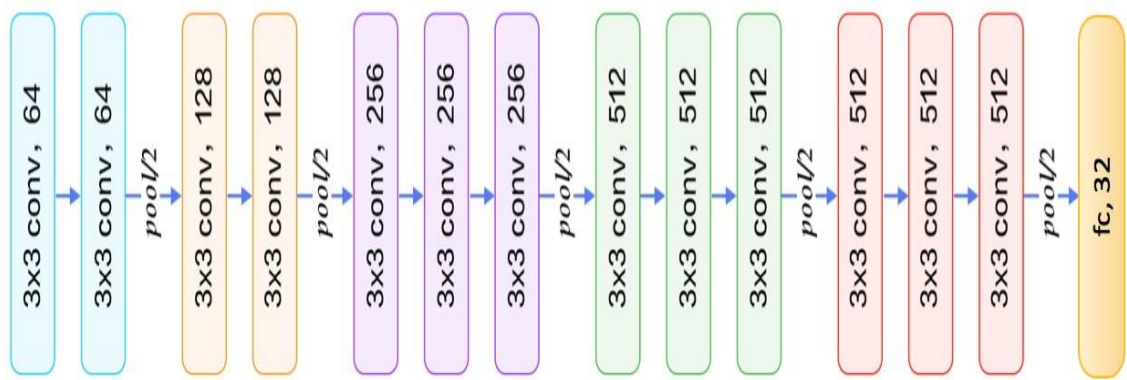

**Figure 9:** Architecture of the best model

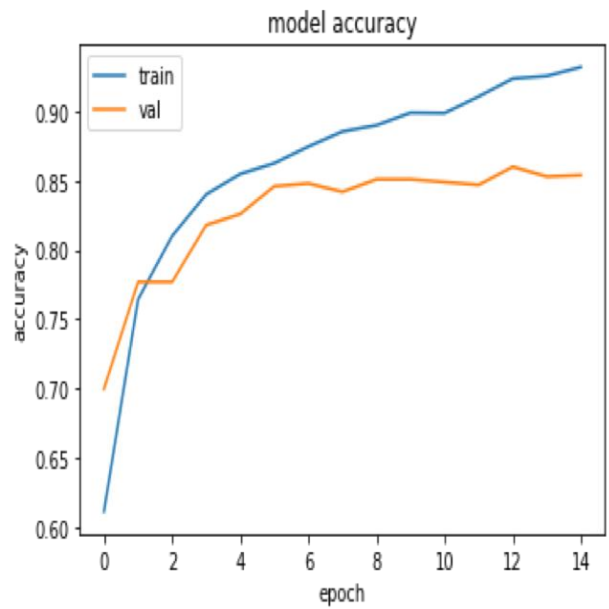 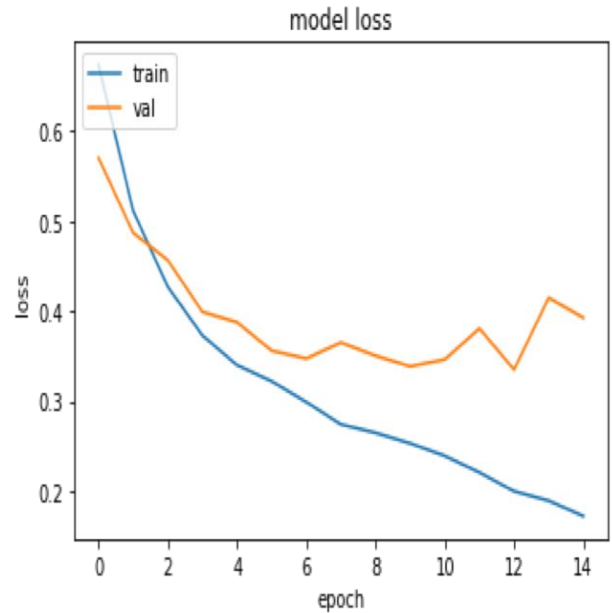

(a) Loss function vs epoch  (b) Accuracy vs epoch

**Figure 10:** Training the best model

After testing three architectures, it is seen that Modified VGG16 outperforms both ResNet50 and MobileNet. MobileNet produces lower accuracy. Modified VGG16 model outperforms ResNet50 model. It can be assumed that this might be due to the features of the images and fixation of ResNet50's loss function on a higher local minimum.

It is also observed that pre-training model provides a better performance compared to the data augmentation. Because the initial weights might have enabled the model in order to find a better local minimum of the loss function during the gradient descent process. It can also be effective to run the model with extra resources and data augmentation for more epochs as the convergence is slow owing to large size of the data.

## 4.2 ROC Analysis

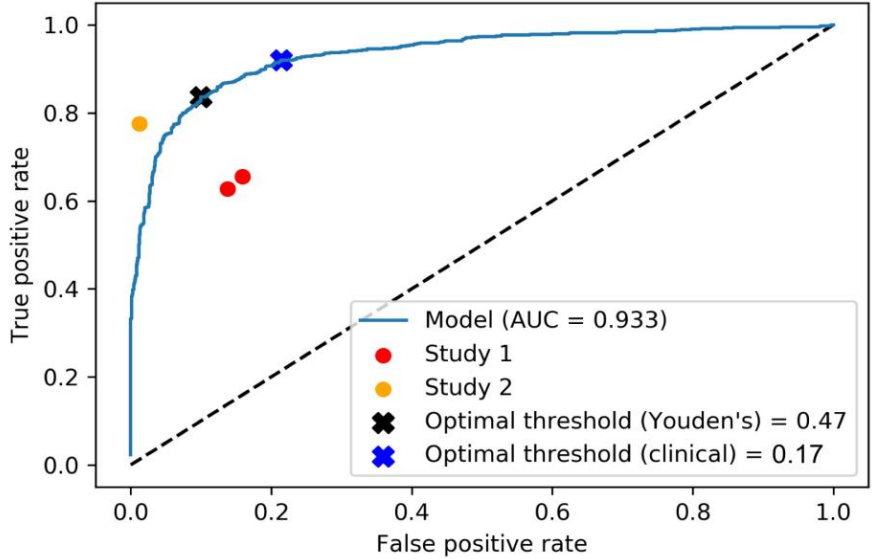

**Figure 11:** ROC curve of the final model

From Figure 11 it is showed that the final modified proposed model has an AUC value of 93.3%. This is better than our benchmark of an AUC value of 88% from Shen et al [17]. Additionally, our model also outperforms radiologists in order to classify the mammograms as pathological or not. The benchmark of the first study on 12 radiologists on 312 cases with a sensitivity of 65.5% and a specificity of 84.1% Rafferty et al. [18] was clearly surpassed by our model. For study 2 the physicians did not only use mammography but also other diagnostics, which could be a reason for the better results.

After finalizing the algorithm, we estimate the mathematically optimal mode for the algorithm is, i.e. what the best threshold for the algorithm to declare a mammogram as either pathological or non-pathological. We compute the Youden's J statistic as follows:

$$J = maximum sensitivity(c) + specificity(c) - 1 \qquad (1)$$

In different words, this threshold minimizes the error rate of false positive and false negative, taking both as equally important. However, as written in chapter 3.4. from a clinical perspective, reducing false negatives is more important than false positives. Thus, we decided to weigh reducing false negatives twice as important as false positive and calculated the optimal threshold maximizing the cost function = 0.66 * true positive rate + 0.33 * (1-false negative rate). The clinically optimal threshold is 0.17, enabling us to further increase the false positive rate (thereby decreasing the false negative rate) by 10% while increase the false positive rate by 15%.

Conclusively, this model with its well performing accuracy as well as the estimated clinically relevant threshold would be well suited to sufficiently reduce errors, especially false negatives, in the clinical setting.

Several data mining algorithm is applied for cancer detection and classification [22-24] using the dataset as a csv file but disease detection and classification using image dataset is a challenging task. In order to classify images into multiple categories such as benign, malignant [25], and normal, our focus is to implement binary classification. This is because classifying a case as normal with higher confidence is more clinically relevant and immediately applicable than multinomial classification. The strength of the paper is the balance between breadth and depth in the scope. Testing various transfer learning models can enable us to recognize the best model for the task. On the other hand, there is some shortcomings with more time consuming and computing resources. So, we have tried to satisfactory tune our proposed models better, trying different hyper-parameters and constructing our own network.

## 4.3 Comparison among the others work of accuracy

Our proposed method provides a higher accuracy than other method proposed in Table 5 for another transfer learning method in same [7-9, 11, 12] and different [4] dataset. It can be explored from Table 6 that the architectures in [7- 9], [12], [11] provide an accuracy of 87.20%, 79.00%, 87.50%, 81.25%, 79.20% and 71% respectively, whereas our proposed model provides the accuracy of 88.3%.

**Table 6:** Comparative analysis with other methods

| References | Authors | Accuracy |
|---|---|---|
| [7] | Rakhlin et al. | 87.20% |
| [8] | Kwok | 79.00% |
| [9] | Vang et al. | 87.50% |
| [12] | Nawaz et al. | 81.25% |
| [11] | Sarmiento et al. | 79.20% |
| [4] | Bharati et al. | 71% |
| **Proposed architecture** | | **88.3%** |

## 5 Conclusion

The best performing architecture is a modified VGG16 network which has been pre-trained on the ImageNet with an accuracy of 88.3% and the value of AUC is 93.3%. The clinical analysis shows that recall or sensitivity should be highlighted over specificity in the case of breast cancer. Thus, a clinically classification threshold is chosen, which is much lower than the mathematically threshold value. This algorithm will help to considerably decrease the false negative cases of mammograms. This will also increase the chances of 5-year survival.

# 6 Future Work

There are different approaches we will like to follow in the future. Instead of a binary classification, it will be interesting to make a categorical classification based on the BI-RADS scores. But in this situation, masses and calcification have to keep merged. We will also need additional, very detailed data set containing not only the BI-RADS scores but also other medical explanations.

Additionally, it will be interesting to integrate additional features into our algorithm. The tissue density of the woman plays a critical role in the breast cancer assessment. Obtaining this and adding it as a feature could potentially increase the accuracy of the algorithm significantly.

**Acknowledgement:** The authors are grateful to those who have participated in this research work.

**Conflict of Interest:** The author declares that they have no conflict of interest

**Ethical approval:** The data used for this research work is collected from DDSM (Digital Database for Screening Mammography) database and authors do not violate the ethical statement.